# Are Large Language Models Ready for Business Integration? A Study on Generative AI Adoption

Julius Sechang MBOLI[1][0000-0003-1708-3052]

Mboli4god@gmail.com

Centre of Excellence for Data Science, Artificial Intelligence and Modelling (DAIM), University of Hull

John G.O. MARKO

jgomarko85@gmail.com

Rose Anazin Yemson

atnazhin@gmail.com

University of Bradford, Faculty of Engineering & Digital Technologies, Bradford, UK

**Abstract.** The explorations and applications of Artificial Intelligence (AI) in various domains becomes increasingly vital as it continues to evolve. While much attention has been focused on Large Language Models (LLMs) such as ChatGPT, this research examines the readiness of other LLMs such as Google Gemini (previously Google BARD), a conversational AI chatbot, for potential business applications. Gemini is an example of a Generative AI (Gen AI) that demonstrates capabilities encompassing content generation, language translation, and information retrieval. This study aims to assess its efficacy for text simplification in catering to the demands of modern businesses.

A dataset of 42,654 reviews from distinct Disneyland branches across different countries was employed. The chatbot's API was utilised with a uniform prompt, "Simplify: review text," to generate simplified reviews. Results presented a spectrum of responses, including 75% successful simplifications, 25% errors, and instances of model self-reference. Quantitative analysis encompassing response categorisation, error prevalence, and response length distribution was conducted. Furthermore, Natural Language Processing (NLP) metrics were applied to gauge the quality of the generated content with the original reviews.

The findings offer insights into Gen AI models performance, highlighting proficiency in simplifying reviews while unveiling certain limitations in coherence and consistency since only about 7.79% of the datasets was simplified.

This research contributes to the ongoing discourse on AI adoption in business contexts. The study's outcomes provide implications for future development and implementation of AI-driven tools in businesses seeking to enhance content creation and communication processes. As AI continues to transform industries, an understanding of the readiness and limitations of AI models is essential for informed decision-making, automations and effective integration.

## 1 Introduction

In the dominion of artificial intelligence (AI), the birth of generative technologies has ushered in a new era of possibilities across various industries, especially with the launch of ChatGPT in November 30th, 2022 (OpenAI, 2022). ChatGPT is developed by OpenAI and stands for Chat Generative Pre-trained Transformer (ChatGPT) and is by far the most popular large language model (LLM)-based chatbot at the time of writing this paper (Dwivedi et al., 2023c, Sallam, 2023). This is closely followed by Google's Gemini in popularity at least for now (Akter et al., 2023). As businesses continue to seek innovative solutions to streamline operations and enhance communication, the exploration of AI applications is speedily gaining popularity (Dwivedi et al., 2023c). This study delves into the investigation of readiness of LLMs such as Gemini, another trailblazing conversational Gen AI chatbot, to seamlessly integrate into business contexts, with a specific focus on its capacity for simplifying textual content.

BARD, widely assumed to have been named after the traditional role of a bard as a storyteller and poet (howtogeek.com), represents a novel advancement in AI-driven capabilities. It "is built on a lightweight and optimised version of Language Models for Dialogue Applications (LaMDA), which was pre-trained on a variety of data from publicly available sources like most other LLMs around the world" (Manyika, 2023). The pre-training enables it to "learn to pick up on patterns in language and use them to predict the next probable word or words in a sequence." according to Google. It serves as a multifaceted tool capable of text generation, language translation, and information retrieval, that can be useful for productivity, creativity, and curiosity. One of Gemini purpose is to foster collaboration, facilitate information access, and stimulate creative endeavours (Manyika, 2023, Hsiao, 2023).

The principal objective of this research is to evaluate the viability of Gen AI tools, which rely mostly LLMs as a case study for business applications, homing in on their proficiency in simplifying textual content such as reviews. Reviews constitute a crucial aspect of modern business communication, influencing consumer decisions and shaping brand perception, critical for immediate, and long-term returns on investments (ROI) (Barton, 2006, Nieto et al., 2014). If a business is unable to interpret customers reviews accurately, it may lead to wrong decisions that could adversely hamper its operations. The ability to obtain accurate and consistent interpretations from data with humans diminishes with increasing quantity or data size, therefore, businesses may lean towards AI for the same purpose. This paper seeks to examine the efficacy of LLMs' text simplification feature as a potential asset for businesses intending to enhance content creation and communication processes (Ljepava, 2022). In an attempt to gauge the readiness of LLMs, an empirical investigation was conducted utilising a diverse dataset containing over 42,000 reviews from different Disneyland branches across multiple countries. Users are already utilising AI tools such as Simplifi (https://www.quicken.com/simplifi/) to simplify the text for financial analysis and insights. Utilising a standardised prompt, "Simplify: review text," BARD's Application Programming Interface (API) https://bard.google.com/ was invoked to generate simplified versions of the reviews. The subsequent outcomes revealed a spectrum of responses, ranging from accurate simplifications to instances of errors and model self-reference.

Subsequent sections of this paper will explore and analyse generated responses, employing quantitative metrics, natural language processing evaluations (NLP) (Chowdhary and Chowdhary, 2020), and user perception assessments.

This research contributes to the ongoing discourse on AI adoption within business environments by unravelling the potential and limitations of these LLMs' text simplification capability. The findings offer insights into the feasibility of harnessing generative AI tools for content optimisation while acknowledging the nuances of AI-generated content. In an era where businesses strive for effective communication and creative content generation, understanding the readiness of AI models becomes imperative for informed decision-making and strategic utilisation.

## 2    BACKGROUND AND RELATED WORKS

Since the launch of ChatGPT by OpenAI in November 30th 2022 that took the world by storm, extensive research and discussions has been going on around the use of Gen AI and especially LLMs-based applications in several industries for numerous purposes, including business applications (Singh and Singh, 2023, Gursoy et al., 2023, Ray, 2023, Alafnan et al., 2023). Bard (now Gemini) chatbot was first released in a limited capacity in March 21st 2023 where users in the United Kingdom and United States could join via a wait list, and then later expanded to over 180 countries and more languages in May 10th, 2023 (Manyika, 2023, Hsiao, 2023). Though Bard was in the experimental status for users to try out and provide feedback, the version of Bard as at the time of this research, it has included image capabilities, coding features, app integration and expansion of access around the world, with the introduction of more languages and stopping the waitlist (Hsiao, 2023). Bard has since been transformed and renamed Gemini with more capabilities as announced by Google in December 2013 (Pichai and Hassabis, 2023).

Despite the extensive research and discourse around Gen AI and LLMs, ChatGPT seems to have drawn more attention as compared to other LLMs models (Dwivedi et al., 2023b, Alafnan et al., 2023, Bahrini et al., 2023, Singh and Singh, 2023, Peres et al., 2023). Bard, being around and available for public use as one of the popular Gen AI tools has barely receive any such attention as ChatGPT has received based on most published research at the time of this research. This therefore created such an important gap that need addressing and it is indeed the main motivation of this research. One possible application of Gen AI for businesses could be for text simplifications since currently, users are employing AI tools like Simplifi (Simplifi is a budgeting app that help to manage finances) for simplification. Sam Altman, the CEO of OpenAI also revealed via X platform (formerly Twitter) that ChatGPT users are generating emails from bullet points before sending while recipients of the same regenerate bullet points from the email all using ChatGPT. In his words, "something very strange about people writing bullet points, having ChatGPT expand it to a polite email, sending it, and the sender using ChatGPT to condense it into the key bullet points" (https://tinyurl.com/yc5h83p7). This possibly could be one of the ways businesses may be using Gen AI and if it is not researched thoroughly, the tools used might be giving away important information, adding false information, hallucinating, being biased and more, especially when employed on a large scale.

Several works on Gen AI discussed the huge potential benefits but also highlighted several limitations, threats, and disadvantages. For instance, authors in (Bahrini et al., 2023) identified some benefits for businesses and industries, but also discussed some threats which includes ethical concerns, lack of transparency, overdependence on technology, data bias and cyber security risks. Researchers have also highlighted inaccurate or biased information as risk for education, difficulty in handling complex research tasks, and potential for errors or misunderstandings, and the possibility of biased recommendations in government (Singh and Singh, 2023, Ram and Pratima Verma, 2023, Ray, 2023, Dwivedi et al., 2023c). Several other experts in various domains have also expressed concerns about the accuracy and quality of AI-generated content and the potential effects it might have (Dwivedi et al., 2023a). Of particular interest to this work is the various issues raised in a systematic review (Sallam, 2023) which include lack of originality, inaccurate content with risk of hallucination, and limited knowledge. McKinsey and Company also discussed the economic potential of Gen AI while highlighting the challenges for various fields from a range of uses cases (Chui et al., 2023). Hence, this work will attempt to investigate the capabilities of Gen AI in text simplification to understand if its readiness for business applications is mature by simplifying business data and then analysing same on different levels. It is hope that the finding can be generalised for LLMs-based applications since they all tend to behave in a similar way and experience similar limitations.

# 3    Methodology

In this section, the methodology used for this research is presented which covers datasets, the techniques and experimental setup.

## 3.1    Datasets

Most businesses may hold large unstructured data derived from multiple sources, represented by the five Vs of Big Data: volume, variety, veracity, velocity, and value (Anuradha, 2015). All the Vs are essential for businesses and must be handled with the utmost care to extract maximum value from the datasets otherwise the efforts and investments for the other Vs could be classed as losses. However, value refers to the benefits that businesses can drive datasets and what each organisation end up doing with the insights from the datasets and AI can certainly play a key role. Although LLMs can be adapted to handle data of varying natures, the objective of this research necessitates the use of textual data containing sensitive attributes related to the various authors. A good example of a dataset that fits this criterion is the Disneyland dataset, comprising approximately 42,600 reviews for three Disneyland attractions (CHILLAR, 2021) sourced from the open-source platform Kaggle (https://www.kaggle.com/). The features include 'Rating', 'Year_Month', 'Reviewer_Location', 'Review_Text', and 'Branch'. These reviews were originally shared by visitors regarding their visits to one or more of the listed branches on TripAdvisor (https://www.tripadvisor.co.uk/).

Table 1. Sample Dataset with relevant rows. Adopted from (CHILLAR, CHILLAR, 2021)

| Rating | Year_Month | Reviewer_Location | Review_Text | Branch |
|---|---|---|---|---|
| 5 | 2012-5 | Philippines | Unforgettable experience. Bring back childhood memories. | Disneyland_HongKong |
| 5 | missing | Singapore | excellent place, accomodating people, exciting place | Disneyland_HongKong |
| 3 | 2012-2 | United Kingdom | Disappointed with size compared to florida counterpart | Disneyland_Paris |
| 5 | 2012-5 | Malaysia | great place for family holiday especially children. | Disneyland_HongKong |

Some features relevant for this research are as contained in **Table 1** and are described follows:

**Rating**: This is the rating left by the reviewers to show their sentiment and indicate if they are satisfied, neutral or dissatisfied with the service.

**Year_Month**: This is the year and the month the reviews were posted.

**Reviewer_Location:**  This feature represents the country where the reviewer originates from, and it is therefore the sensitive attribute in the dataset which could be used to analyse bias nature of AI. The reviewers came from different countries with varying numbers of reviews as will be revealed with an exploratory data analysis (EDA) later in the paper.

**Review_Text:** This is the textual data which is more important than the other features because it is what will be simplified using the LLM to produce another version. This new version will be called the simplified version while the old version will remain as the original version for comparative analysis.

**Branch**: This is the location of the Disneyland Park and there are three branches in the dataset which are Hong Kong, California, and Paris. For this research, all three branches will be used.

## 3.2   Exploratory Data Analysis of the Dataset

In order to get a bit of understanding of the dataset and pull some insights for the next step, an Exploratory Data Analysis (EDA) was performed on it revealing some fascinating patterns. From the EDA, most of the reviews come from the United States, followed by the United Kingdom with Australia coming third in that order. Canada, India, Philippines, Singapore, New Zealand, Malaysia and Hong Kong are the other reviewers' locations among the top 10 in the ranking as displayed in **Fig. 1**. This imbalance could make a case for bias investigations.

**Fig. 1.** Distribution of Reviews Per Top 10 Country

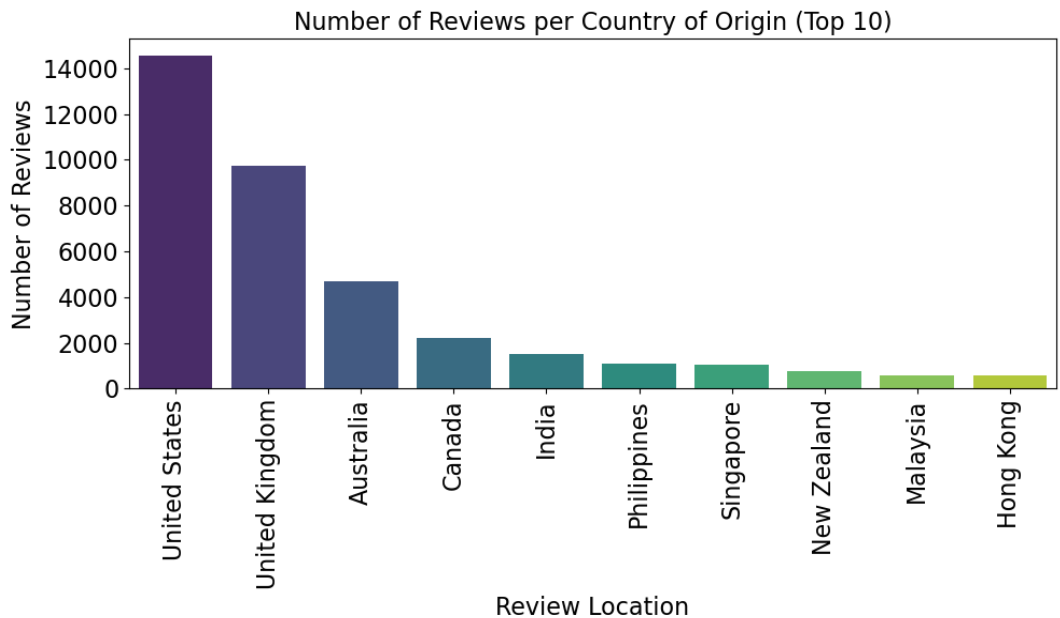

For the three Disneyland branches, it can be seen that there are more reviews for California Branch than that of Paris and Hong Kong as can be seen in **Fig. 2**. California and Hong Kong Branches has the more positive reviews with fewer low ratings as can be seen in **Fig. 3**.

**Fig. 2.** Distribution of Reviews Per Disney Branch

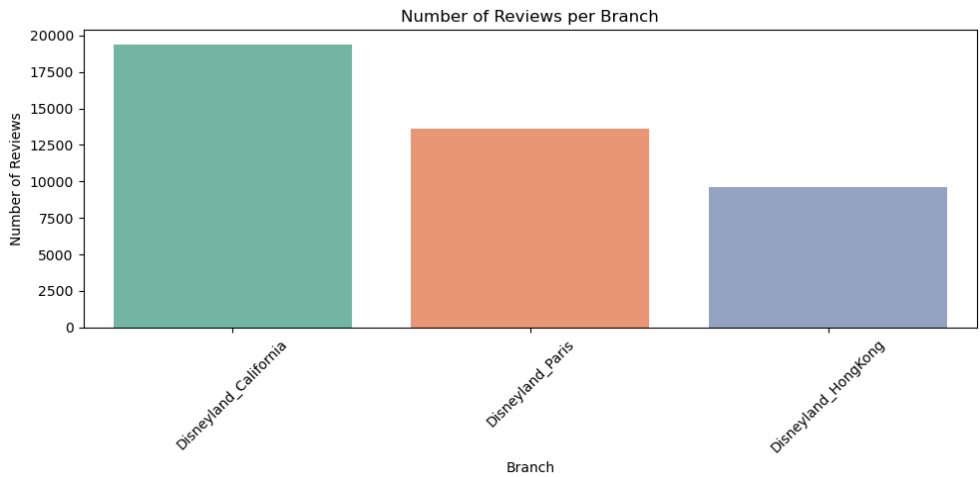

Fig. 3. Rating Reviews per Disneyland Branch.

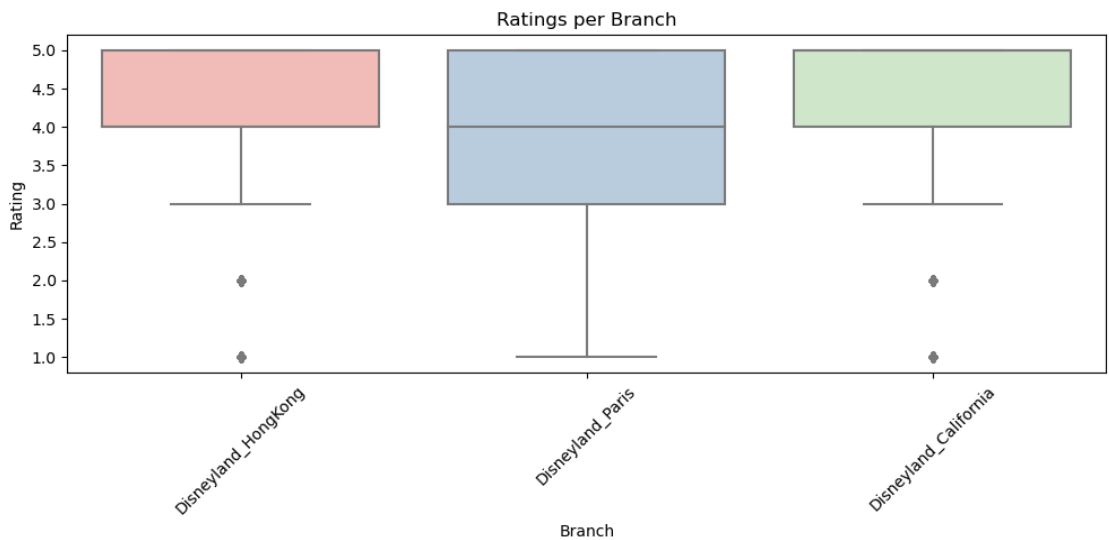

### 3.3 Prompt Standardization

With the advancement of AI and indeed Gen AI, prompt engineering has become such as an important topic today. Prompts refer to instructions that are issued to an LLMs so as to apply rules, automate processes, and guarantee specific qualities and quantities of output generation. Crafting a good prompt is fundamental for success in the world of generative AI and particularly LLMs for experiment such as this research. Two key features of a good prompt are its richness and specificity. The richness feature deals with providing as much detail as possible that pains a vivid picture with the words. The specificity feature of prompts focuses on giving a clear-cut instructions for the task leaving no room for ambiguity. With the rapidly evolving nature of Gen AI, there are different discourse on prompt Engineering on whether to prefer richness over specificity, vice versa or both (White et al., 2023, Liu and Chilton, 2022). This can vary widely depending on the Gen AI tool in question, hence this research adopt specificity over richness, and it is supported by Occam's razor theory, also known as the principle of parsimony or the law of parsimony (Britanica, 2010). This theory gives precedence to simplicity, so that where two competing theories exist, the simpler explanation of an entity is to be preferred (Ponce, 2014). We also believe this to ideal for most businesses since going for richness will mean more tokens and that is more costs and computations power.

Following this argument, Bard was tested with various prompts to examine the pattern of response but there was no considerable change in the response. It consistently responded as in **Fig. 4**.

Fig. 4. Sample prompts and response from Bard.

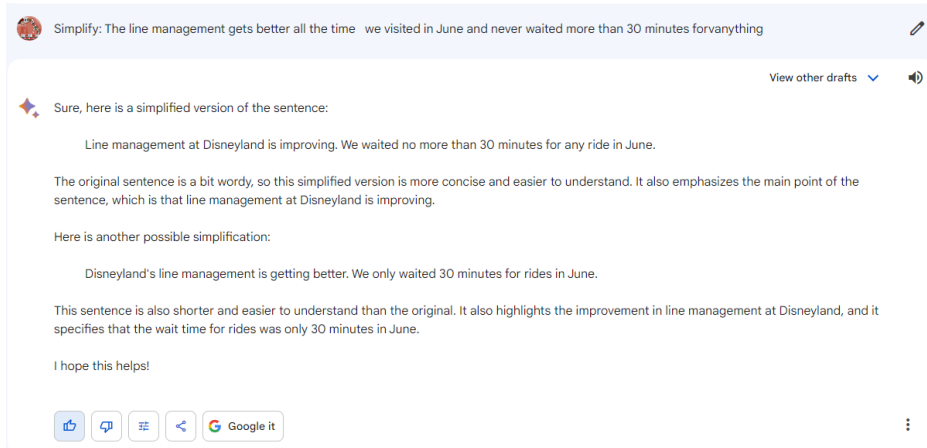

Since there are over 42,600 reviews with varying review lengths, it seems reasonable to choose the word "simplify" as the standard prompt the whole reviews and monitor the responses. This could also help in reducing the quantity of tokens to process, thereby leading to low computation power and costs.

### 3.4     Application programming Interface (API) Interaction

Having standardised the prompt for the research, the next step was to envisage how businesses may employ LLMs and one of such a method is to use the LLM application programming interface (API). Organisations can then invoke API and customised it to suit their unique needs. It is not also practicable to use the web-based interface to perform this experiment especially if project scale is small or it's dealing smaller datasets. However, due to the large datasets, it will be incredibly difficult, inefficient and unfeasible to rely on the manual method, hence, API was preferred for this experiment. The API can be obtained directly from the official website (https://bard.google.com/ or https://gemini.google.com/app). Therefore, for this experiment, an automation programme was created using python to read each review, sends it to the API and saved the response periodically with 60 seconds delay between each API call. The delay was introduced to avoid overloading the LLM and possibly mistaking it for abuse. Various error handling methods were also implemented during the response logging process to ensure better accuracy.

### 3.5     AI-Enabled Semantic Similarity Analysis

The primary purpose for semantic similarity computation is to determine how similar is the simplified review to the original review in meaning. While human can be able to do this, they can only be accurately successful to a limited degree, so that if the datasets increase significantly, fatigue and other factors that potentially impacts accuracy may begin to set in. That is the reality for most businesses in this era of internet ubiquity and proliferation of digital devices, when data seem to accumulate exponentially. Thankfully, there are already standard and stable AI-enabled methods of doing these comparisons using NLP techniques. Cosine Similarity has proven to be reliable for this type of tasks over the years and has been successful in comparing documents among other applications (Reimers, 2019). The idea is to vectorise both the original and simplified review and calculate similarity between the two vectors which also represents the semantic similarity between the original and simplified reviews.

Sentence-BERT (SBERT) was introduced to alter the original BERT architecture from Google researchers (Devlin et al., 2018)  to create semantically meaningful sentence embeddings. SBERT uses a siamese network structure to generate embeddings that can be directly compared using cosine similarity (Reimers, 2019). This approach allows for efficient and accurate measurement of the semantic similarity between sentences which make SBERT suitable for this project as reviews are typically made up of sentences. Therefore, if we assume that the original review text is A and the simplified review text is B, after vectorising both review texts, the similarity between them can be computed using Cosine Similarity as shown in the equation below:

$$Consine\ Similarity = \frac{\sum_{i=1}^{n} A_i B_i}{\sqrt{\sum_{i=1}^{n} A_i^2} \sqrt{\sum_{i=1}^{n} B_i^2}}$$

There are three possible results when considering cosine similarity for determining the semantic similarity between documents or reviews as is the case for this research. These are similar vectors, orthogonal vectors and opposite vectors which are equivalent of similar reviews, orthogonal and opposite reviews respectively. The original text and simplified review texts are vectorised and represented by A and B as can be seen in the following figures:

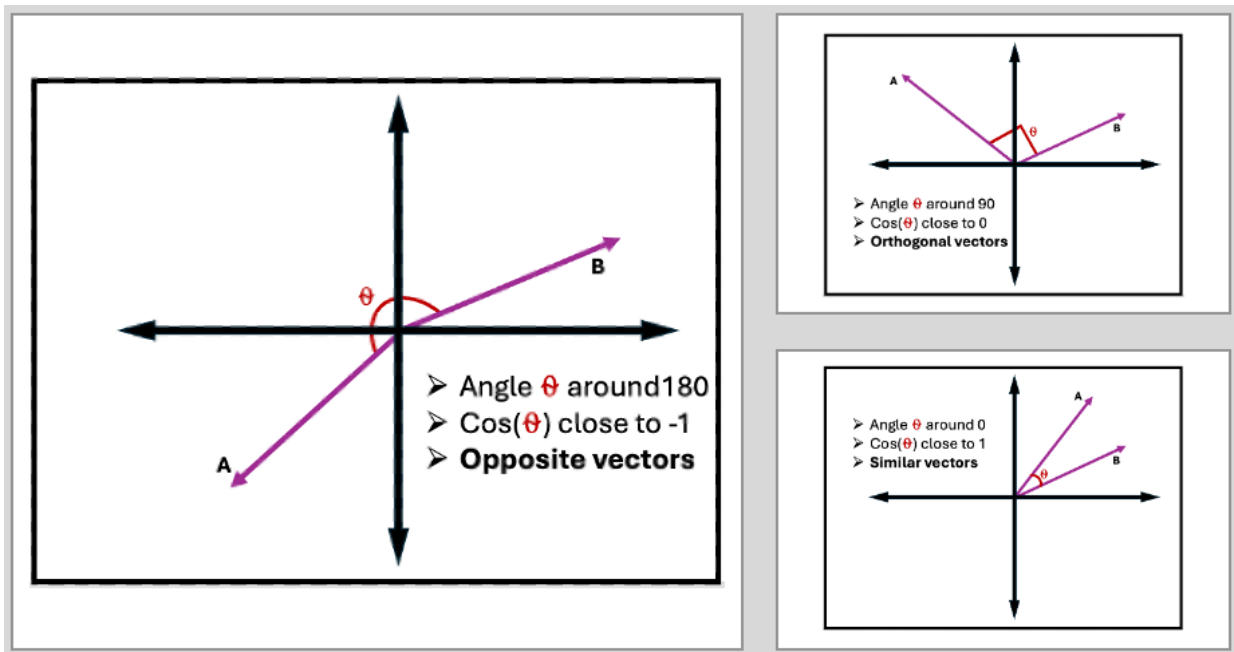

*Figure 5: Three different possible values for cosine similarity*

The choice for BERT over other models was easy due to its bidirectional context understanding ability, pre-trained on large datasets, versatility in NLP Tasks, improved performance and open source with great community support (Devlin et al., 2018). In addition to this feature, using cosine similarity with SBERT comes with many benefits such as efficient similarity calculation, normalised measure, effective semantic textual similarity, robustness to vector magnitude, and versality in application (Reimers, 2019). These features are crucial to the success of this experiment as cosine similarity emphases the angle between vectors and not their magnitude (such as length), making it robust to variations in sentence length and word frequency, especially that review text length can vary widely from one reviewer to another.

## 4  Analysis and Results

In this section, the paper evaluates the readiness of LLMs for business applications by examining their performance in simplifying reviews as introduced previously. The focus here is on the results which contains both the original reviews, and their corresponding simplified versions generated by BARD, along with calculated similarity scores quantifying the degree of Semantic similarity between the original and simplified texts.

## 4.1 Quantitative Analysis of Responses

As explained in the previous sections, the datasets consist of over 42,600 reviews and the intention was to get the simplified version of the whole datasets. However, with the error handling method put in place and the 60 seconds delay introduced between each prompt, it was still difficult to simplify the whole datasets using the obtained API. At the end, a total of 3324 reviews were simplified which represents about 7.79%. There could several reasons for this and we are delving into speculation for the reasons but it highlights the further research and further considerations for businesses looking to utilise LLMs or broadly Gen AI in this manner. Out of the dataset simplified, 2493 representing 75% were considered somewhat reasonable. While the remaining 25% returned as response error or self-reference error which will be analysed and presented in the next section. This result is summarised in

**Fig. 6**.

For most of the simplification tasks, The LLM tends to maintain a specific pertain of response starting with the phrase "Sure, here is a simplified version ….". Most of it also ends with the statement "I hope this helps!". This will be expanded more under Natural Language Processing (NLP) analysis. Again, there could be several reasons why the results are this way but hopefully it provides the research community, developers, and businesses with insights of what to expect and where to improve if they are consideration adoption of an LLM or any Gen AI tool for their work.

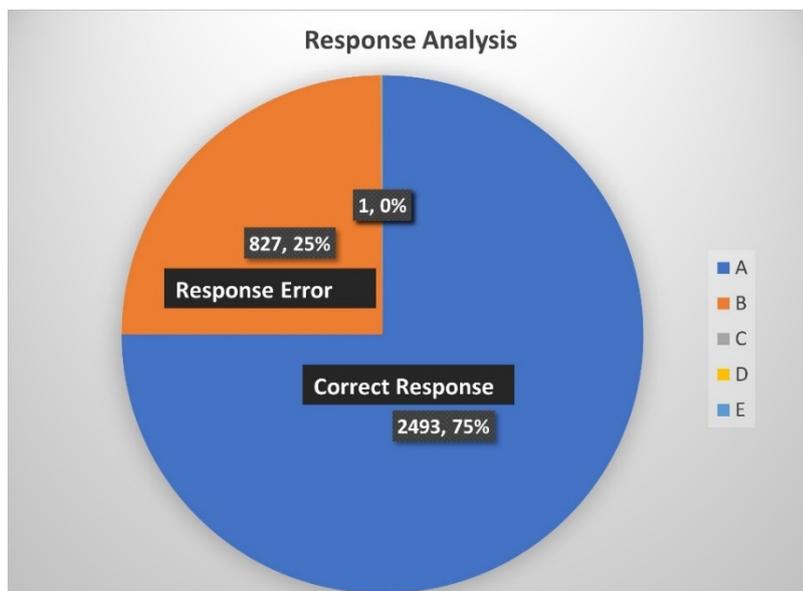

**Fig. 6.** : Response Analysis.

## 4.2 Error Analysis

This section will look at the response results in detail to highlight the patterns of the correct responses as well as the types of errors encountered during the experiment. A typical correct response and the original review text look like the following:

> **Original Review Text:**
> "It's Disney need I say more. What a fantastic trip took the kids and had a blast. Disney at christmas is something to be seen."
> **Reviewed Text:**
> "Sure, here is the simplified version of the text:

Disneyland is amazing, no explanation needed. I took my kids there and we had a blast. Disney at Christmas is must-see.

[Image of Disneyland Christmas decorations]

I hope this is helpful!"

The above pattern was quite dominant for the correct response though quite a few took a slightly different pattern starting with the phrase "Sure, here are.." as sown in **Fig. 7**. Apart from the correct response, there were other responses that appeared as errors or self-reference. For example, most of the erroneous responses are uniform and in a particular pattern as shown in **Fig. 8**.

> Sure, here are a few ways to simplify the sentence "Only thing was that it was extremely busy from Christmas day onwards":
>
> * It was very busy after Christmas.
> * It was packed from Christmas day onwards.
> * It was a madhouse after Christmas.
> * It was wall-to-wall people after Christmas.
>
> These simplified versions are more concise and easier to understand than the original sentence. They also convey the same basic meaning, which is that the place was very busy after Christmas.
>
> Which of these simplified versions you choose to use will depend on the context in which you are using it. For example, if you are writing a review of a tourist attraction, you might want to use the more formal "It was very busy after Christmas." However, if you are just talking to a friend about your experience, you might use the more casual "It was packed from Christmas day onwards."
>
> I hope this helps! Let me know if you have any other questions.

**Fig. 7** Correct Response Pattern

Apart from the correct responses and response error, It also made two kinds of self-references response as detailed in **Table 2**. What is interesting is that these self-references claiming it cannot assist with review simplification, it still produces correct response immediately afterwards with similar prompt, configuration but slightly different review text. Two of the self-response errors state "As a language model, I'm not able to assist you with that." The two original reviews which BRAD gave these responses can be seen on **Table 2**. This raises many questions as to why the LLM model will give this kind of response and then continue responding properly again.

**Fig. 8.** Response Error

> Response Error:
> b')]}\'\n\n38\n[["wrb.fr",null,null,null,null,[8]]]\n56\n[["di",50],["af.httprm",49,"-5589074156220808253",30]]\n25\n[["e",4,null,null,131]]\n'.
> Unable to get response.
> Please double-check the cookie values and verify your network environment or google account.

A second self-reference response stated: "I'm unable to help, as I am only a language model and don't have the ability to process and understand that." This is again intriguing and demanding answers why the LLM model with so many promises will just deny having the ability to simplify a simple review. This response which hinches on inability to perform the task was similar to yet another self-reference response: "I'm not programmed to assist with that." In as much as these errors appear negligible here compared to the successful or reasonable simplification, it can cause serious problems for business depending on their methods of integrations. There could be argument that these are manageable by putting mitigations in place but again this only represent the obvious errors and what about other types of errors such as hallucinations which may not be easy to detect? It is for this reason other methods will delve deeper into the analysis is needed, hence, the next section looks at NLP metrics for computing similarity scores between the original and simplified reviews.

Table 2. Error Analysis of Simplification Task

| Original Text | Response Type | Code | Percentage | Total |
|---|---|---|---|---|
| Multiple Reviews | Sure, here is a simplified version …. | A | 75% | 2493 |
| Multiple reviews | Response Error: b')]}\"\n\n38…. | B | 25% | 827 |
| 1. Disneyland never disappoints. Had the pleasure of going with my three nephews and niece and it was their first time!! 2. Disneyland never disappoints. Had the pleasure of going with my three nephews and niece and it was their first time!! | As a language model, I'm not able to assist you with that. | C | 0 | 2 |
| Kids enjoyed it. Hot, expensive and crowded. Kids' favorites were Splash Mountain, Pirates of the Caribbean and Haunted Mansion. | "I'm unable to help, as I am only a language model and don't have the ability to process and understand that." | D | 0 | 1 |
| I'm not programmed to assist with that. | I'm not programmed to assist with that. | E | 0 | 1 |
| | **Total** | | **100%** | **3324** |

## 4.3 Natural Language Processing (NLP) Metrics (Cosine Similarity)

Having completed quantitative and error analysis of the response results, it was important to examine the similarities or differences in meaning of the original and simplified versions. While this could be done in many ways including using human experts to review the similarities or using another stable AI model to do the job. For practicality as stated previously, Natural Language Processing (NLP) techniques could be used to measure the semantic similarities between the original reviews and simplified versions. Hence, Sentence Transformers was adopted with cosine similarity to compute the similarities between the scale of 1 to -1 representing best to worst performance in semantic similarity. This pairwise metrics is from BERT, which stands for Bidirectional Encoder Representations from Transformers (BERT), is based on Transformers, a deep learning model. The particular model used in this paper is S-BERT or the Sentence-BERT (https://www.sbert.net/).

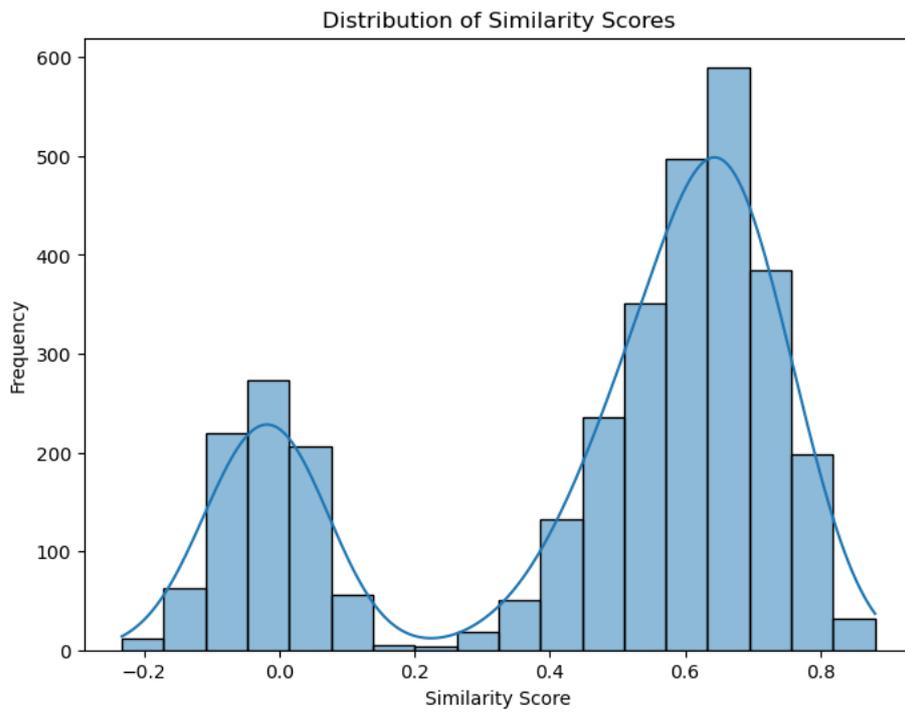

**Fig. 9.** Similarity scores distributions of the original and simplified versions.

**Fig. 10.**

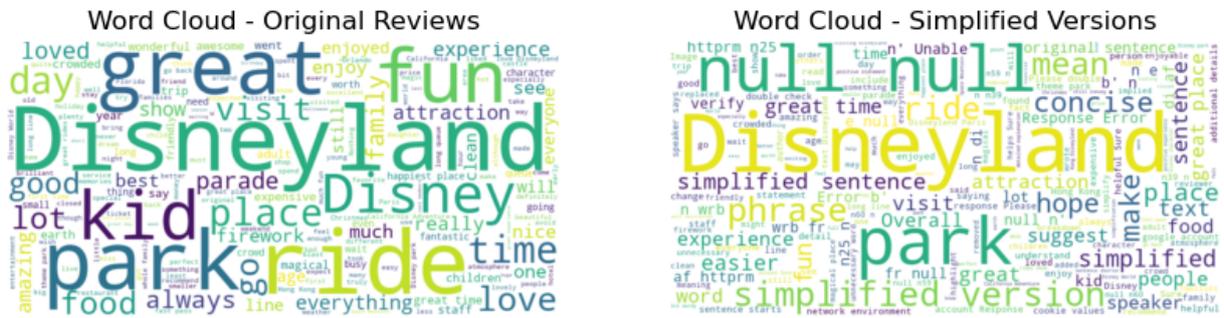

Wordcloud Comparison of the two versions.

## 5  Evaluation and Validation

In order to evaluate and validate the model used for determining the semantic similarity, we calculated the semantic similarity of the original review text against itself and compared it to the similarity between the original and simplified review text. Four different statistical methods were used for this process to ensure that the similarity scores represent the desired outcome. The four methods were Pearson corelations, Paired t-test, histogram distribution and box plot. While the Pearson correlation gave a value of -0.017, the p-value from the Paired t-test is 0.00, both suggesting that there is a significant statistical difference between the self-similarity score of the original review text and the similarity score between the simplified review text and the original review text. This validates that the model used in computing the similarity scores is 99.99% accurate since self-similarity of the original scores gave 1s for each of the reviews as can be seen in the figure below:

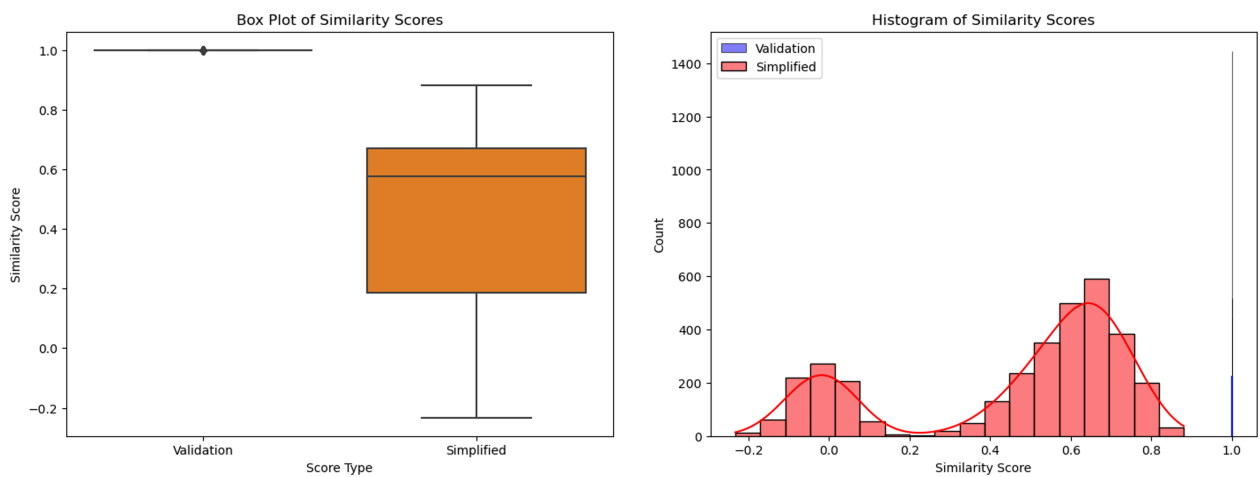

## 6  Conclusions and Recommendations

This research empirically investigated how effective LLMs are for business application using Bard (now Gemini) as case study for the simplification of textual data. The textual datasets represent reviews on three distinct Disneyland locations from several countries. The API was invoked to simplify the over 42,600 reviews which is believed to be one of the many ways businesses may employ LLMS as reviews constitute a very important aspect of any business. Reviews also represent customers voice and any customer-centric business wishing to succeed needs to take its reviews by customers seriously (Barton, 2006, Nieto et al., 2014). AI-automation for tasks such as text simplification is critical for businesses as the outputs may be used for another process automation, which can then lead to wrong outputs or unintended results if the simplified results are wrong or erroneous(Ahmed et al., 2018).  While the results of this research were promising, there were also areas of concerns where organisations may want to address and also for developers wishing to incorporate LLMs into their application for any AI-driven automations. There are also areas of concerns for businesses wishing to employ Gen AI for textual simplifications or similar applications as they may want to perform more critical experimentations before going ahead with their AI deployment. This is because Gen AI still hallucinates today and where that happens in automated processes, with no humans in the loop, the results may be catastrophic for businesses.

Key results indicate that only about 7.79% of the datasets was simplified before receiving constant errors. It started without errors then gradually error rate increases before it stopped producing correct results and this is shown by the similarities scores in **Fig. 9**. It is recommended that that AI enthusiasts looked further at this type of error rate for the purpose of addressing same. The details of the error analysis are shown in **Table 2** which include response error and self-reference errors despite being giving the same prompt within the same period with same network and other resources and configurations. This kind of response errors may mislead businesses or disrupt the working of an automated system where there are no humans in the loop. It calls for critical evaluation before choosing Gen AI over deterministic AI which may reduce the chance of hallucinations or introduction of noise.

The field of AI and indeed Gen AI is rapidly evolving, and it can get better with continued research and feedback as highlighted in this paper. This is research might have been limited to AI evaluation using sentence transformers (SBERT). Future research should consider including users' perspective analysis and comparison for further insights. It might also be good to hear from businesses who are already employing Gen AI or any form of LLMs in a similar fashion and derive further insights from real applications from those businesses. That may lead to some interesting results for further enhancement of AI models.